\documentclass[a4paper,twocolumn,fleqn,10pt]{article}
\usepackage{custom}
\usepackage[authoryear]{natbib}
\usepackage{siunitx}

\title{A cornucopia of null results: A statistical analysis of \\ fireballs reported to the American Meteor Society}
\author[1]{Althea Moorhead}
\affil[1]{NASA Meteoroid Environment Office, Marshall Space Flight Center EV44, Huntsville, Alabama, USA}
\date{}

\begin{document}
%\linenumbers

\twocolumn[
  \begin{@twocolumnfalse}
    \maketitle
    \begin{abstract}

        In March 2026, the American Meteor Society announced that a ``surge'' of large fireballs had been reported to their website in the first quarter of the year, and that these fireballs had certain characteristics (radiant clustering and reports of delayed sound).
        We find this data set to be an excellent use case for Poisson regression, which, in our opinion, is underutilized in meteor astronomy. This report serves as a brief primer on Poisson regression and related statistical techniques as well as an analysis of AMS fireball counts. We find that the number of events reported in early 2026 is in line with the overall pattern of activity. We also find little evidence of the ``February fireballs'' phenomenon.
    
    \end{abstract}
  \end{@twocolumnfalse}
  \vspace{\baselineskip}
]

%%%%%%%%%%%%%%%%%%%%%%%%%%%%%%%%%%%%%%%%%%%%%%%%%%%%%%%%%%%%%%%%

% Main text
\section{Introduction}

The American Meteor Society (AMS) collects visual fireball reports through their website; a web application helps users record their location and viewing angle \citep{hankey14}. These reports are used to estimate meteor trajectories by the AMS as well as by other groups, including the NASA Meteoroid Environment Office \citep{moser17}. On March 25, 2026, the American Meteor Society reported that they had collected an unusually high number of meteor reports in the first quarter of 2026 \citep{hankey26}. The post included the following claims (paraphrased here):
\begin{enumerate}[itemsep=0em]
    \item A ``surge'' of fireballs with 25+ reports occurred during the first quarter of 2026 (tested in \S\ref{sec:rev}).
    \item The surge is stronger at higher reporting thresholds (tested in \S\ref{sec:interact}).
    \item The increase is most noticeable in March (tested in \S\ref{sec:month}).
    \item There were an unprecedented number of ``major'' fireballs in mid-March (not tested; see \S\ref{sec:untested}).
    \item A large fraction of fireballs with 100+ reports during the first quarter of 2026 were reported to produce delayed sound (tested in \S\ref{sec:sound}).
    \item A large number of events with long durations (${\ge 4}$\,s) were reported during the first quarter of 2026 (not tested; see \S\ref{sec:untested}).
    \item Fireballs with 25+ reports have radiants that are unusually clustered in two regions of the sky compared to the previous four years (tested in \S\ref{sec:rad}).
\end{enumerate}
The post was later updated to include additional data and a modified analysis \citep{hankey26b}. Both versions solicit independent analyses, which we are happy to provide in this paper.

\cite{hankey26} also mentions a piece of meteor astronomy lore sometimes called the ``February fireballs.'' According to several news interviews with Dr.\ Peter Brown, amateur astronomers ``noticed an increase in the number of deep-penetrating, bright meteors during February back in the 1960s and `70s'' \citep{phillips12,space12}. According to \cite{lapaz49}, February has been ``notable for
meteoric phenomena of unusual intensity'' since the 1920s. However, peer-reviewed studies of fireball rates generally do not find that rates are elevated in February \citep[e.g.,][]{rendtel89,halliday96}. We will not resolve this debate here using a second-hand set of visual reports with unknown biases (e.g., weather patterns). We test only whether there is evidence of an unusual number of fireballs \emph{reported to the AMS} in a given period.

In order to test the AMS's claims, we will need to look for statistically significant variations in the number of fireballs reported during specific intervals and with certain characteristics. We will assume that the mean fireball rate is a possible function of time and reporting threshold, and that the number of reported fireballs follows a Poisson distribution about this mean. These assumptions suggest the use of Poisson regression.

Poisson regression is used to model count data in a wide variety of fields ranging from traffic analysis to ecology \citep[examples include][]{miaou92,verhoef07}. Like linear regression, it is a form of machine learning \citep[although we prefer the term ``statistical learning'';][]{james21}.
Poisson regression is little used in meteor astronomy, despite the fact that many studies in this field analyze meteor or photon counts; \cite{ozerov24} provides a rare example of its use. We therefore provide a brief primer on Poisson regression in \S\ref{sec:poisson} before applying it to the AMS data. Sections~\ref{sec:resids} and \ref{sec:multiple} also outline our approach to residual inspection and performing multiple comparisons. Finally, \S\ref{sec:sce} discusses the importance of using Sun-centered ecliptic (SCE) coordinates when analyzing the distribution of meteor radiants. We use these tools in \S\ref{sec:application} to test five of the seven enumerated claims (two claims were not testable). We find little evidence in the AMS data of any unusual activity in early 2026.

\section{Methods}
\label{sec:methods}

This section describes the statistical tools used in our analysis. If the reader is already familiar with Poisson regression, quantile residuals, the Bonferroni correction for multiple comparisons, and SCE spherical coordinates, we suggest skipping ahead to \S\ref{sec:application}.

\subsection{Poisson regression}
\label{sec:poisson}

The reader is almost certainly familiar with linear regression performed via ordinary least squares (OLS), in which one fits for the slope and intercept by minimizing the sum of the squared residuals. The reader may not be aware, however, that linear regression is a member of a larger family of generalized linear models (GLMs). In this section, we provide a brief overview of GLMs, with an emphasis on Poisson regression. 

Suppose that we have a data set with one or more predictor variables (${x_1, ..., x_p}$) and one response variable ($y$). Generalized linear models assume that the expected value of the response variable (denoted $\mathrm{E}(y)$ or $\mu_y$) can be related to the predictor variables as follows:
\begin{align}
    g (\mu_y) &= \beta_0 + \sum_{j=1}^{p} \beta_j x_j \, ,
\end{align}
where $g$ is known as the link function.
This model describes only the expected mean of the response; the actual response values are assumed to follow some probability distribution function (PDF) whose mean is $\mu_y$. For some combinations of PDF and link function, it is possible to derive an analytic expression for the most likely values of $\beta$. These combinations tend to be implemented in statistical software packages; we list a few in Table~\ref{tab:combos}.

\begin{table}[b]
    \centering\small
    \begin{tabular}{lcccc} \hline \hline
        link function & 
         \multicolumn{4}{c}{PDF or GLM ``family''} \\ 
         & binomial & Poisson & gamma & normal \\ \hline
        identity &
            & \checkmark & \checkmark & \checkmark \\
        inverse &
            && \checkmark & \checkmark \\
        square root &
            & \checkmark && \\
        log &
            \checkmark & \checkmark & \checkmark & \checkmark \\
        logit &
            \checkmark &&& \\
        probit &
            \checkmark &&& \\
    \end{tabular}
    \caption{A few of the most common GLM ``families'' and link functions implemented in R (glm in base R) and Python (GLM in the statsmodels package). A check mark indicates that a particular combination of family and link function has been implemented. The ``logit'' and ``probit'' functions are the quantile functions of the logistic and normal distributions, respectively. The normal distribution may be referred to as ``Gaussian'' in these software packages.}
    \label{tab:combos}
\end{table}

When the link function is the identity function and the response variable is assumed to follow a normal distribution with constant variance about $\mu_y$, the calculated $\beta$ values are identical to those obtained via OLS. However, we are interested in measuring counts, and will turn our attention to the Poisson GLM family -- that is, to Poisson regression.

Poisson regression can be used to model the rate at which events occur, if these events satisfy certain assumptions. Events must occur one at a time and must be independent \citep[i.e., the occurrence of one event does not affect the likelihood of a second event occurring; see, e.g.,][]{ott16}. The rate can vary with time, location, or other external factors, but the expected number of events within an interval is the integral of the rate over that interval.

It is not always necessary to use Poisson regression to analyze count data; when the expected count is large, a Poisson distribution resembles a normal distribution with a mean of $\mu_y$ and a standard deviation of $\sqrt{\mu_y}$. One can then perform weighted least squares, or unweighted OLS after regularizing the data \citep{johnson05,ohara10}; see \cite{moorhead26} for an example application. These approaches fail when rates are low and counts of zero appear in the data. In contrast, empty bins pose no problems for Poisson regression.

The variance of a Poisson distribution is equal to its mean; if the data have a larger variance than this, it can indicate either a lack of fit or overdispersion. Overdispersion could indicate that the data do not truly follow a Poisson distribution, or that there are are hidden (unmeasured) factors influencing the count rates. Excess variance can be diagnosed using the residual deviance: if data are Poisson-distributed and consistent with the model, the residual deviance will be comparable to the number of residual degrees of freedom (the number of observations minus the number of fitted parameters). If the residual deviance is much larger, this can indicate lack of fit or overdispersion. One should also examine the fit residuals (\S\ref{sec:resids}) to determine whether there is an obvious lack of fit. If the data appear to fit the model but overdispersion is present, quasi-Poisson regression can be used. Quasi-Poisson regression and Poisson regression will produce the same fit parameters, but quasi-Poisson regression will account for the observed dispersion when calculating fit uncertainties.

\begin{figure}%\footnotesize
    \includegraphics[width=\linewidth,trim={0 0.55in 0 0},clip]{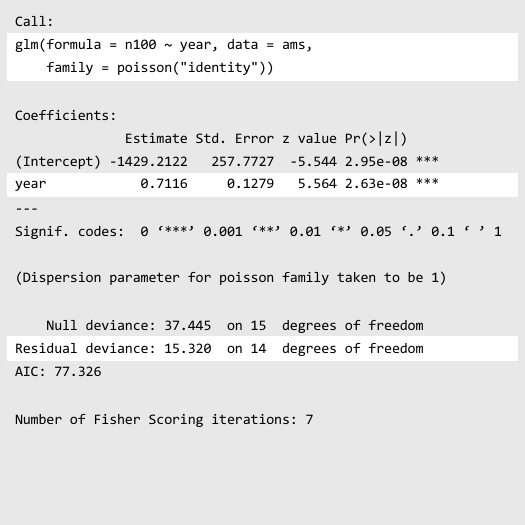}

    \caption{Example summary from a GLM run in R, with the most useful lines highlighted.}
    \label{fig:output}
\end{figure}

Figure~\ref{fig:output} presents sample output from a simple GLM run in R. A model has been fit to a data set named ``ams'' that contains a predictor variable named ``year'' and a response variable named ``n100'' (the number of events with at least 100 reports). We have chosen a Poisson model with an identity link function. From the coefficients table, we see that the number of events with 100+ reports grows by an average of 0.7116 per year. The p-value gives the probability that the slope differs from zero under the model assumptions; a value of ${p = 2.63 \times 10^{-8}}$ tells us that ``year'' is a statistically significant predictor for the number of events. We find the residual deviance and degrees of freedom towards the bottom of the output summary. 

For a thorough discussion of GLMs, we refer the reader to chapter~4 of \cite{agresti13}. For guidance in implementing GLMs in Python or R, we refer the reader to section~4.6 of \cite{james21} or \cite{james23}, respectively, both of which are available for free on the authors' website.\footnote{\url{https://www.statlearning.com/}}

\subsection{Residual inspection}
\label{sec:resids}

When fitting any model, it is important to examine the residuals for signs of lack of fit or violations of model assumptions. When working with response data that are normally distributed, one can examine the Pearson residuals. These can be calculated as follows:
\begin{align}
    z_{P,\,i} &= \frac{y_i - \mu_i}{\sigma_i}
\end{align}
where $\sigma_i$ is the standard deviation of ${y_i - \mu_i}$, and we use $i$ to enumerate the observations. If the response data are consistent with the model and satisfy the assumptions made in least-squares linear regression, the Pearson residuals will 
\begin{itemize}[itemsep=0em]
    \item show no dependence on any of the predictor variables ($x_j$) or the predicted response ($\mu$), 
    \item show no dependence on adjacent residual values,
    \item display a constant variance (homoskedasticity), and
    \item follow a standard normal distribution.
\end{itemize}

The Pearson residuals can also be used to identify outliers. If the model assumptions are valid and the model fits the data, we expect 95\% of residuals to lie between $-2$ and 2, and 99.7\% between $-3$ and 3. One point with ${z_P = 5}$, for example, in a small data set would be considered an outlier.

A Poisson distribution resembles a normal distribution when its mean is large. (The definition of ``large'' is subjective, but is approximately 10--20.) Thus, for large counts, we can still opt to examine the Pearson residuals:
\begin{align}
    z_{P,\,i} &= \frac{y_i - \mu_i}{\sqrt{\mu_i}} \, . \label{eq:normapprox}
\end{align}
When the counts are low, however, the normal approximation breaks down. One alternative is to use quantile residuals:
\begin{align}
    z_{q,\,i} &= F_\text{normal}^{-1} \left( \, F_\text{Poisson} ( y_i \, ; \, \mu_i ) \, \right) \, ,
\end{align}
where $F_\text{normal}^{-1}$ is the inverse cumulative distribution function (CDF) of a standard normal distribution, and $F_\text{Poisson}$ is the CDF of a Poisson distribution with mean $\mu_i$. Thus, quantile residuals map the residuals to a standard normal distribution, which is useful to viewers accustomed to assessing Pearson residuals. When applied to discrete data, randomization is introduced to compensate for granularity \citep{dunn96}.

\subsection{Multiple comparisons and outlier identification}
\label{sec:multiple}

It is customary to select a significance level, usually denoted $\alpha$, prior to conducting any significance testing, with a common choice being ${\alpha = 0.05}$. Test results are then considered statistically significant if the p-value is less than $\alpha$; the interpretation is that there is only a 5\% chance of obtaining evidence against the null hypothesis (no effect) that is at least as strong as that observed. 

However, it is very common to perform multiple significance tests in a single study. If ten tests are performed, the probability of obtaining a p-value below 0.05 by pure chance -- in the absence of the hypothesized effect(s) -- is 40\%. This is known as the \emph{multiple comparisons problem}.

One simple way of handling multiple comparisons is to define a per-test significance threshold:
\begin{align}
    \alpha_\text{test} &= 1 - (1 - \alpha)^{1/n_\text{test}} 
    \simeq \alpha/n_\text{test} \, ~ \text{ for } \alpha \lesssim 0.05 \, ,
\end{align}
where $n_\text{test}$ is the number of significance tests performed and $\alpha$ is the overall significance threshold. This is known as the Bonferroni correction, and, at a minimum, should be applied when conducting multiple tests on a single data set. In this author's opinion, $n_\text{test}$ should reflect the total number of significance tests performed in the course of completing the study, whether or not they are mentioned in the final publication. 

The Bonferroni correction can also be applied to outlier identification. If a data set contains only 20 observations, a single observation with a residual of ${|z| > 5}$ is unexpected. However, if the data set contains $10^7$ observations, there is a 99.7\% probability that at least one observation will have a residual of ${|z| > 5}$. If one is specifically interested in testing whether outliers are present in the data, one can compare the residuals to:
\begin{align}
    z_\text{test} &= \pm F_\text{normal}^{-1} \left( 1 - \frac{1}{2} \frac{\alpha_\text{test}}{n} \right) \,
\end{align}
where $n$ is the number of observations in the data set. The factor of one-half accounts for the fact that outlier identification is generally a two-tailed test.

The Bonferroni correction is the simplest method for accounting for multiple comparisons. There are, however, other approaches; we readers to \S7.5 of \cite{agresti13} for more information.

\subsection{Sun-centered ecliptic (SCE) coordinates}
\label{sec:sce}

\begin{figure}[!b]
    \centering
    \includegraphics[width=0.6\linewidth]{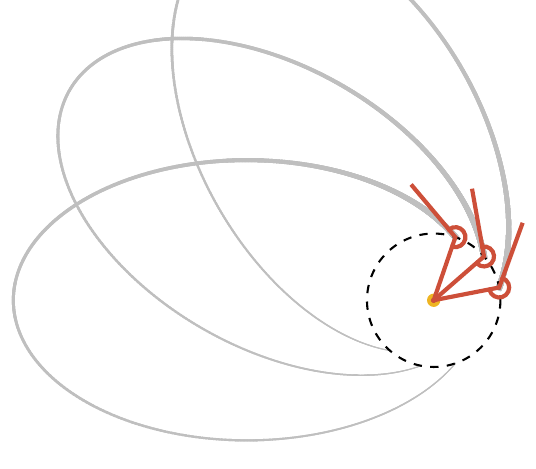}
    \caption{In this diagram, the Earth's orbit (dashed black circle) is intersected by three meteoroid orbits (gray ellipses) whose orbital elements are the same except for the longitude of ascending node. The Sun-Earth-radiant angle (thick red lines) -- and thus the SCE radiant -- is the same in all three cases.}
    \label{fig:sce}
    \vspace{\baselineskip}

    \includegraphics{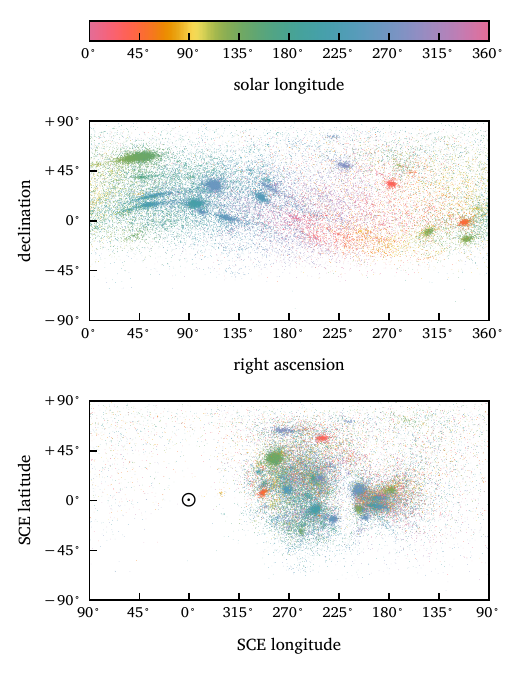}
    \caption{Radiants of meteors observed by the NASA All Sky Fireball Network in both equatorial (top) and Sun-centered ecliptic (SCE; bottom) coordinates. Points are color-coded by solar longitude (i.e., time of year). The position of the Sun is marked in the bottom panel. These data are not part of our analysis and are included here only to illustrate the utility of SCE coordinates.}
    \label{fig:allsky}
\end{figure}

Meteoroid orbits evolve over time in response to radiative forces -- which is less important for large meteoroids -- and the gravitational influence of planets. Planetary perturbations can cause meteoroid orbits to precess in both argument of pericenter (apsidal precession) and longitude of ascending node \citep[nodal precession; see][for Taurid precession rates]{asher93}. Unlike apsidal precession, nodal precession does not affect an orbit's ability to intersect the Earth: it simply changes the time of year when the intersection occurs. Nodal precession also preserves the Sun-Earth-radiant angle (see Fig.\,\ref{fig:sce}). It is therefore useful to examine radiants in a non-inertial coordinate system that is aligned with the ecliptic and the Sun-Earth vector: we call these coordinates Sun-centered ecliptic (SCE) coordinates.

The use of SCE coordinates is not in itself a statistical technique. It can, however, be considered a form of dimension reduction, as it allows us to substitute two variables (SCE longitude and latitude) for three (right ascension, declination, and solar longitude) for certain applications. 

The utility of SCE coordinates can be seen in Fig.\,\ref{fig:allsky}, which presents the radiants of meteors observed by the NASA All Sky Fireball Network \citep{kingery20} in both equatorial and SCE coordinates. It is customary to reverse the longitude axis and center the data on an SCE longitude of $270^\circ$, which is aligned with the Earth's direction of motion or ``apex'' direction. We have also marked the position of the Sun at ($0^\circ$,~$0^\circ$).

Meteor showers generally resemble streaks in equatorial coordinates and clumps in SCE coordinates (see Fig.\,\ref{fig:allsky}). The lack of daytime meteor observations is also apparent in SCE coordinates, as the Sun's position in this coordinate system is always ${(0^\circ, 0^\circ)}$. The Sun's angular position is only indirectly apparent in equatorial coordinates, as the range of radiant positions that are visible at night shifts over the course of the year, introducing a relationship between right ascension and solar longitude.  
Finally, it is possible to see the variations in background radiant density due to the sporadic sources in SCE coordinates. For these reasons, any analysis of the distribution of meteor radiants should be conducted using SCE coordinates.

\section{Application to AMS reports}
\label{sec:application}

The \cite{hankey26} post provides some, but not all, of the data used in this section. First-quarter event counts per year are provided in a table within the post; these counts are used in \S\ref{sec:cum}. The post also links to a data file containing radiants for fireball events with at least 25 reports in the first quarter of the year between 2021 and 2026 (inclusive).\footnote{\href{https://fireball.amsmeteors.org/members/imo_view/browse_events}{https://fireball.amsmeteors.org/members/imo\_view/browse\_events}} This radiant data is used in \S\ref{sec:rad}.

We also downloaded HTML data tables from the AMS website. These tables were accessed using the ``Events'' button and provide the date and report counts for individual events. These data are used to bin events by report count (\S\ref{sec:binned}) and by month (\S\ref{sec:month}). Event data dating back to 1980 is available through the website, but we limit our analysis to 2011--2026. We made this choice based on the statement in \cite{hankey14} that the AMS reporting system was upgraded in late 2010 to a version that is better able to locate the user's location and viewing angle. We did not include earlier events in our analysis at any point. 

The AMS also has an application programming interface (API) for data downloads. We found, however, that the data available through the API was not consistent with that available through the website or with \cite{hankey26b}: the number of reports tended to be lower, and radiants often differed by large angles (almost 180$^\circ$ in one case). We therefore opted to use the data from the website in order to be consistent with \cite{hankey26b}.

The use of Poisson regression (\S\ref{sec:cum}--\ref{sec:month}) requires that the data -- the number of reported fireball events in different time intervals -- be independent. Variations in fireball rates due to observing conditions, seasonal variations, or meteor showers do not violate this requirement. Independence could be violated by clusters of nearly simultaneous meteors \citep{koten24} or by erroneously splitting reports of a single event into multiple events. We did not find any simultaneous events in the AMS data, nor did we find an excess of near-simultaneous events.

We have included a supplementary Markdown file showing how we performed our analysis. This file contains our calculation of meteor radiants and solar longitudes in Python, and all of our statistical analyses, which were performed in R, along with their output. Readers can access the data from the AMS website.

\subsection{Analysis of cumulative counts}
\label{sec:cum}

\begin{figure}[!t]
    \centering
    \includegraphics{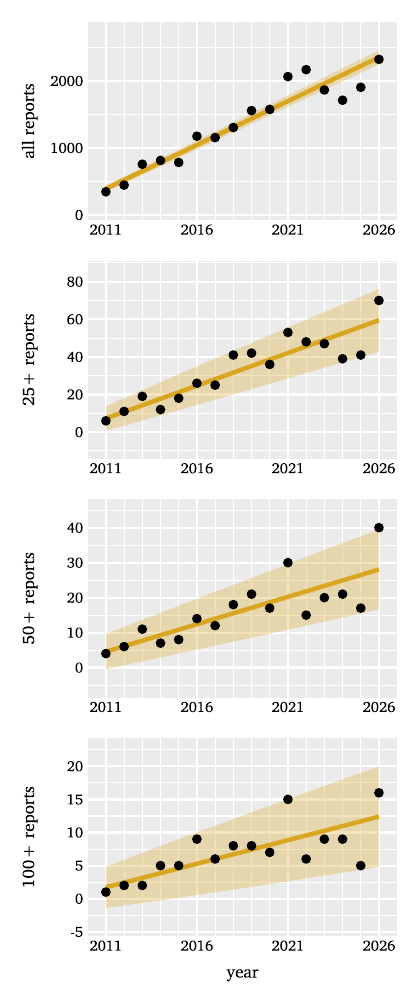}
    \captionof{figure}{The number of fireballs reported to the AMS in the first quarter of each year since 2011: each row corresponds to a different minimum number of reports (1, 25, 50, and 100). The line follows the best-fitting linear relationship between event count and year. We expect 95\% of the data to lie within the shaded region around the best-fitting line. 
    }
    \label{fig:fit}
    \vspace{2\baselineskip}

    \small
    \begin{tabular}{lccC{1.2cm}C{1.1cm}} \hline \hline
        subset & slope & $x$-intercept & 
        residual deviance & residual d.f. \\
        \hline
        all events & 131(9) & 2008.0(0.6) & 337.9 & 14 \\
        25+ reports & 3.5(0.3) & 2008.9(0.6) & 21.5 & 14 \\
        50+ reports & 1.6(0.2) & 2008.1(1.1) & 20.6 & 14 \\
        100+ reports & 0.7(0.1) & 2008.6(1.5) & 15.3 & 14 
    \end{tabular}
    \captionof{table}{Fit coefficients, residual deviance, and residual degrees of freedom (d.f.)\ after performing Poisson regression.}
    \label{tab:dispersion}
\end{figure}

A visual inspection of the AMS count data (see Fig.\,\ref{fig:fit}) reveals that the counts tend to increase over time. The data show larger variations in recent years, which is the expected behavior for Poisson data. We therefore opted to perform Poisson regression on these data using an identity link function and with the year as the only predictor variable. That is, for each reporting threshold, we fit:
\begin{align}
    \mu_{N,\,\text{Q1}} &= \beta_0 + \beta_1 x_\text{yr}
    \label{eq:mod0}
\end{align}
where $x_\text{yr}$ is the integer year and $\mu_{N,\,\text{Q1}}$ is the number of fireballs in the first quarter of the year that meet the reporting threshold.
Note that these event counts are not independent: events with 100 or more reports are present in all four subsets. 

Figure~\ref{fig:fit} shows the best fit to each subset of data. We noticed some commonalities between these fits; for instance, all fits have similar $x$-intercepts (see Table~\ref{tab:dispersion}). We also see that the mean slope of the 50+ report data is a little more than twice that of the 25+ report data, and the 100+ and 50+ report data sets have a similar ratio.

The shaded regions in Fig.\,\ref{fig:fit} correspond to ${z_P = \pm 2}$, where $z_P$ is the Pearson residual calculated using the normal approximation to a Poisson distribution (see eq.\,\ref{eq:normapprox}); around 95\% of the data should lie within the shaded region if our assumptions are correct. Events with 25+, 50+, and 100+ reports satisfy this expectation. This is also reflected in the residual deviance (Table~\ref{tab:dispersion}): events with 25+, 50+, and 100+ reports have a residual deviance comparable to the number of residual degrees of freedom, and thus pass the dispersion check. 

When we perform Poisson regression on all events, including single-report events, we see fluctuations in the count data that are much larger than expected. Table~\ref{tab:dispersion} shows that the variance is inflated by a factor of 20. This suggests there are other sources of variation in the number of low-report-count events. This could be because these events tend to be dimmer, and therefore are less likely to be seen in poor weather. It may also very well be that single-report events are more likely to be misidentified non-meteor events such as aircraft or satellites.

\subsection{Analysis of binned counts}
\label{sec:binned}

As noted in \S\ref{sec:cum}, all four reporting thresholds appear to share a common $x$ intercept, and the slope amplitude appears to vary with reporting threshold in a predictable way when that threshold is at least 25 reports. This suggests that we may be able to fit the data with a single, global model that varies with year ($x_\text{yr}$) and reporting threshold ($x_\text{rep}$):
\begin{align}
    \mu_{N,\,\text{Q1}} &= A_0 \, (x_\text{yr} - 2009) \, x_\text{rep}^{-\beta_\text{rep}} \, , \label{eq:myfunc}
\end{align}
where $A_0$ is a constant of proportionality. We cannot fit this relationship using the data shown in Fig.\,\ref{fig:fit}, as Poisson regression assumes that the response data (counts) are independent, and this is not the case when events with 100+ reports contribute to the counts at all four reporting thresholds.

\clearpage
We prepare a valid data set by dividing events into disjoint subsets: for instance, we replace the ``25+ reports'' category with a ``25--49 reports'' category. This redivision assures that the resulting counts are independent. We also discard events with less than 25 reports, given the overdispersion exhibited by these data, and add a ``200--399 reports'' category so as not to discard too many big events. 
Finally, we include the latter quarters of the years 2011--2025 in addition to the first quarter of the years 2011--2026.

\subsubsection{Temporal trend}
\label{sec:temporal}

We now re-write eq.\,\ref{eq:myfunc} for fireballs that are binned by the number of reports:
\begin{align}
    \ln \mu_{n,\,Q_j} =~&\beta_{Q_j} 
    + \beta_\text{yr} \ln \left( x_\text{yr}' - 2009 \right)
    + \beta_\text{rep} \ln x_\text{rep} 
    \, , \label{eq:coefs}
\end{align}
where $\mu_{n,\,Q_j}$ is the expected number of fireballs with at least $x_\text{rep}$ reports but less than ${2 x_\text{rep}}$ reports in quarter ${Q_j \in \lbrace \text{Q1}, \, \text{Q2}, \, \text{Q3}, \, \text{Q4} \rbrace}$ and
 $x_\text{yr}'$ is the fractional year: 
\begin{align}
    x_\text{yr}' &= x_\text{yr} + \frac{j - 1}{4} \, , \label{eq:xprime}
\end{align}
where $j$ is the quarter index.
Readers should note the following:
\begin{itemize}
    \item The variable $x_\text{rep}$ has the same definition here as in eq.\,\ref{eq:myfunc}: it is always the lower limit of the report number interval, whether that interval spans [$x_\text{rep}$, $2 x_\text{rep}$) or [$x_\text{rep}$, $\infty$).
    \item We assume that the long-term linear growth in reports is continuous, and therefore fold the quarter index, $j$, into the time-dependent term, after subtracting its mean value during the first quarter (1).
    \item We have incorporated both reporting threshold ($x_\text{rep}$) and fractional year ($x_\text{yr}'$) into our model as quantitative variables.
    \item The quarter, on the other hand, is treated as a categorical variable ($Q_j$) in eq.\,\ref{eq:coefs}, but as a quantitative variable ($j$) in eq.\,\ref{eq:xprime}. 
    \item Fitting for four separate $\beta_{Q_j}$ values allows for seasonal variations in fireball counts.
    \item Equation~\ref{eq:coefs} uses a log link function: ${g(\mu_{n,\,Q_j}) = \ln \mu_{n,\,Q_j}}$.
\end{itemize}

From eq.\,\ref{eq:myfunc}, we expect to find that ${\beta_\text{yr} \approx 1}$. This is confirmed by our fit results: we obtain ${\beta_\text{yr} = 1.02 \pm 0.05}$.  The full set of fit results (and residual deviance) is provided in Table~\ref{tab:fityear}. Readers may note the resemblance between Table~\ref{tab:fityear} and the sample output shown in Fig.\,\ref{fig:output} (aside from the lack of ``z-values''). The estimate and standard error values can be combined to compute confidence intervals for the coefficients: for instance, the 95\% confidence interval for $\beta_\text{Q1}$ is ${4.21 \pm 2 \times 0.19}$.

The p-values indicate whether there is evidence that the coefficient differs from zero; for instance, the small p-value listed for $\beta_\text{yr}$ indicates that the number of fireballs varies linearly with year and that this variation is statistically significant. Many of the p-values in our tables are extremely small ($\lesssim 10^{-100}$); this is not uncommon, especially when the coefficient is obviously non-zero. There is no difference in interpretation, however, between a p-value that is many orders of magnitude smaller than the significance threshold and one that is only slightly smaller than the threshold.

\begin{table}%[!b]
    \small\centering

    \begin{tabular}{l%
        S[table-format=2.2]
        S[table-format=1.2]c}
     & \multicolumn{1}{c}{estimate} 
     & \multicolumn{1}{c}{std.\ error} 
     & \multicolumn{1}{c}{p-value} \\ \hline
    $\beta_\text{Q1}$ & 4.21 & 0.19 & 1.0e$-$108 \\
    $\beta_\text{Q2}$ & 3.74 & 0.19 & 2.0e$-$084  \\
    $\beta_\text{Q3}$ & 4.14 & 0.19 & 4.3e$-$105 \\
    $\beta_\text{Q4}$ & 4.46 & 0.19 & 2.4e$-$123 \\
    $\beta_\text{yr}$ & 1.02 & 0.05 & 9.6e$-$084 \\ 
    $\beta_\text{rep}$ & -1.12 & 0.04 & 3.6e$-$202 \\[5pt]
    \multicolumn{4}{l}{Residual deviance: 284.4} \\
    \multicolumn{4}{l}{Residual deg.\ freedom: 238} \\
    \end{tabular}

    \caption{Coefficient table and residual deviance and degrees of freedom obtained by fitting eq.\,\ref{eq:coefs} to the AMS data.}
    \label{tab:fityear}
\end{table}

\subsubsection{Revised model}
\label{sec:rev}

From this point forward, we will assume ${\beta_\text{yr}=1}$. Our model then becomes
\begin{align}
    \ln \mu_{n,\,Q_j} =~&\beta_{Q_j} 
    + \ln \left( x_\text{yr}' - 2009 \right)
    + \beta_\text{rep} \ln x_\text{rep} 
    \, . \label{eq:by1}
\end{align}
We then fit this revised model and obtain updated values for the remaining parameters (e.g., ${\beta_\text{rep} = -1.12 \pm 0.04}$). The full set of fitted coefficients is presented in Table~\ref{tab:quarterly}. The table also includes the residual deviance, which is larger than the number of residual degrees of freedom but not excessively so.

\begin{table}%[!b]
    \small\centering

    \begin{tabular}{l%
        S[table-format=2.2]
        S[table-format=1.2]c}
     & \multicolumn{1}{c}{estimate} 
     & \multicolumn{1}{c}{std.\ error} 
     & p-value \\ \hline
    $\beta_\text{Q1}$ & 4.26 & 0.14 & 9.5e$-$197 \\
    $\beta_\text{Q2}$ & 3.79 & 0.15 & 1.3e$-$146 \\
    $\beta_\text{Q3}$ & 4.19 & 0.14 & 7.2e$-$188 \\
    $\beta_\text{Q4}$ & 4.51 & 0.14 & 4.3e$-$225 \\
    $\beta_\text{rep}$ & -1.12 & 0.04 & 3.6e$-$202
    \\[5pt]
    \multicolumn{4}{l}{Residual deviance: 284.5} \\
    \multicolumn{4}{l}{Residual deg.\ freedom: 239} \\
    \end{tabular}

    \caption{Coefficient table and residual deviance and degrees of freedom obtained by fitting eq.\,\ref{eq:by1} to the AMS data.}
    \label{tab:quarterly}
\end{table}

\begin{figure*}
    \centering
    \includegraphics{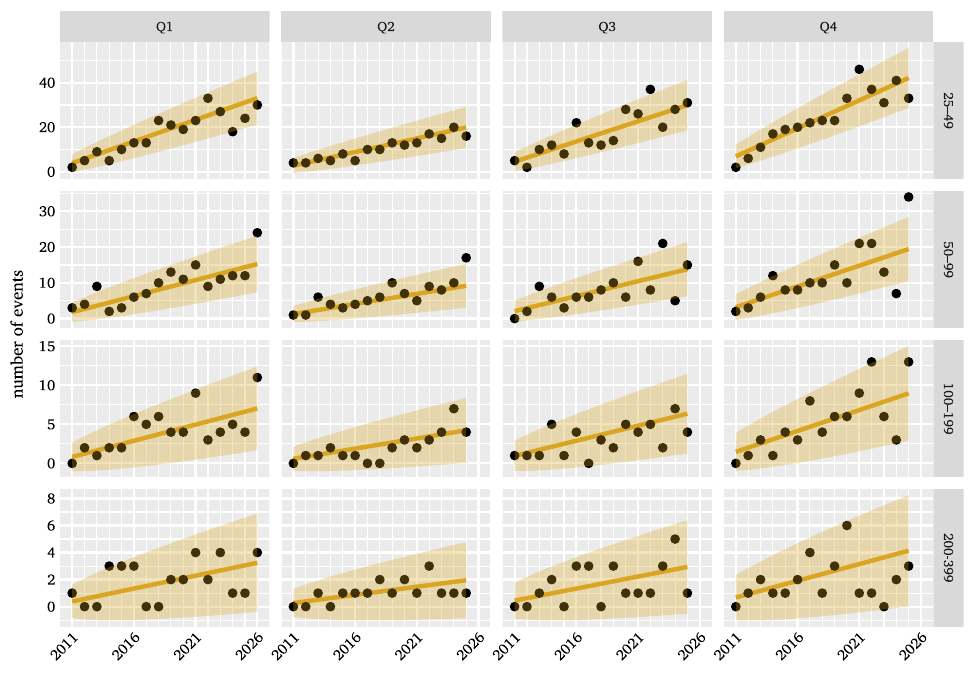}
    \caption{The number of fireballs reported to the AMS every year since 2011: each row corresponds to a different range in the number of reports submitted per fireball, and each column corresponds to a different quarter of the year. The solid line follows our best fit to eq.\,\ref{eq:coefs}, and the shaded region approximates the 95\% prediction interval.}
    \label{fig:grid}
\end{figure*}

Figure~\ref{fig:grid} compares our fit with the data in each quarter and reporting threshold bin. We can use this scatterplot grid to compare activity patterns across years and quarters: for instance, we see no hint of a change in the growth rate around 2016--2018, when, according to the AMS, the reporting platform was considered ``mature.'' One could conceivably interpret the Q1 counts as reaching a plateau prior to 2026, and we presume that this is the reason that \cite{hankey26} compared the 2026 counts with the 2021--2025 average \citep[the updated post used 2018--2025;][]{hankey26b}. This behavior is not consistent across quarters, however, which indicates that the Q1 counts may simply have been lower than average for a few years in a row. Overall, the data appear to be consistent with steady growth in reported events over the time period considered.

\begin{figure}
    \centering
    \includegraphics{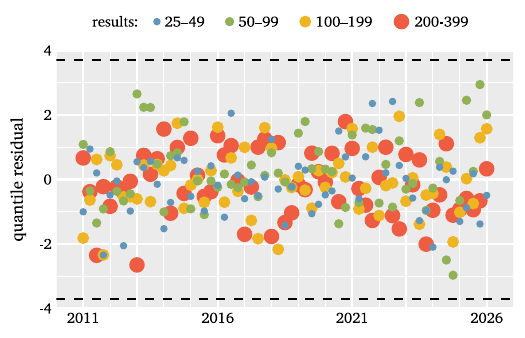}
    \caption{Quantile residuals for the number of fireballs reported to the AMS in each quarter year between the first quarter of 2011 and the first quarter of 2026 (inclusive). Residuals are calculated relative to the best fit to eq.\,\ref{eq:coefs} where $\beta_\text{yr}$ is set to 1. The dashed black lines at ${\pm 3.99}$ correspond to a Bonferroni-adjusted significance threshold of $\alpha_\text{test} = 0.5/244$.  There are no outliers present.}
    \label{fig:resids}
\end{figure}

Figure~\ref{fig:resids} presents the quantile residuals \citep{dunn96} corresponding to Fig.\,\ref{fig:grid} as a function of time and reporting threshold. These residuals form a horizontal band of fairly even thickness, with no apparent grouping of residuals. In other words, there are no obvious signs that our model is inappropriate.

Approximately 20 residuals (recall that there is some randomization in the calculation of quantile residuals) out of 244 have an absolute value greater than 2: this is a little higher than expected (${0.05 \times 244 = 12.2}$). No residuals have absolute values exceeding the Bonferroni-adjusted limit of ${z_\text{test} = 3.71}$, however. We therefore conclude that there is insufficient evidence that the number of reported fireballs in any quarter, at any reporting threshold, and in any year is anomalous (claim \#1).

We end this section by converting our best-fit coefficients into an equation. Because we have subdivided the data by powers of two in $x_\text{rep}$, the expected number of events with at least $x_\text{rep}$ reports is given by the following geometric series:
\begin{align}
    \mu_{N,\,Q} &= \sum_{k=0}^{\infty} e^{\beta_Q} \, ( x_\text{yr}' - 2009 ) \left( 2^k x_\text{rep} \right)^{\beta_\text{rep}} \nonumber \\
    &= e^{\beta_Q} \, ( x_\text{yr}' - 2009 ) \frac{x_\text{rep}^{\beta_\text{rep}}}{1 - 2^{\beta_\text{rep}}}  \, .
    \label{eq:geom}
\end{align}
If we substitute our best-fit parameters into this equation, we obtain the following expected event rates for the four quarters:
\begin{align}
    \mu_{N,\,\text{Q1}} &\approx 131 \, 
        ( x_\text{yr}' - 2009 ) \,
        x_\text{rep}^{-1.12} \, 
        \nonumber \\
    \mu_{N,\,\text{Q2}} &\approx \hphantom{0}83 \, 
        ( x_\text{yr}' - 2009 ) \,
        x_\text{rep}^{-1.12} \, 
        \nonumber \\
    \mu_{N,\,\text{Q3}} &\approx 122 \, 
        ( x_\text{yr}' - 2009 ) \,
        x_\text{rep}^{-1.12} \, 
        \nonumber \\
    \mu_{N,\,\text{Q4}} &\approx 170 \, 
        ( x_\text{yr}' - 2009 ) \, 
        x_\text{rep}^{-1.12} \, .
        \label{eq:geom2}
\end{align}
Equation\,\ref{eq:geom} (and thus eq.\,\ref{eq:geom2}) assumes that our best-fit model applies at all reporting thresholds, including those outside our selected range of 25--399 reports. We therefore perform a quick check by using eq.\,\ref{eq:geom2} to predict the number of events with one or more reports in the first quarter of 2026.  We obtain a prediction of ${\mu_{N,\,\text{Q1}}(x_\text{rep} = 1) = 2227}$ events, which is fairly close to the observed value of 2322 events. This level of agreement also suggests that the overdispersion seen in events with a low reporting threshold may be due to the influence of other factors, rather than to large numbers of reports of non-meteor events.

\subsubsection{Interaction between year and reporting threshold}
\label{sec:interact}

In the previous section, we fit a model to all quarters and checked for outliers. This approach was inspired by our assumption that the AMS would have also commented on an unusually high number of reported fireballs in any previous quarter. In this section, however, we specifically test whether 2026 differs from previous years in either overall activity level ($\beta_\text{Q1}$) or the distribution of the number of reports per fireball ($\beta_\text{rep}$). This allows us to test the claim that the increase in fireballs in the first quarter of 2026 is more pronounced at high reporting thresholds (claim \#2).

We separate 2026 from the rest of the data by introducing a dummy variable, $x_{2026}$, that is equal to one for fireballs observed in 2026 and zero in all other cases. We then allow $\beta_\text{Q1}$ and $\beta_\text{rep}$ to vary with $x_{2026}$, which is accomplished by allowing $x_{2026}$ to ``interact'' with $x_\text{rep}$. The corresponding model is:
\begin{align}
\ln \mu_{n, Q_j} &= \beta_{Q_j}
+ \beta_\text{rep} \ln x_\text{rep}
+ \ln \left( x_\text{yr}' - 2009 \right)
\nonumber \\
& +(\Delta_{\text{Q1}, \, 2026}
+ \Delta_\text{rep, 2026} \ln x_\text{rep} ) \cdot x_{2026} \,.
\label{eq:interact}
\end{align}
We use ``$\Delta$'' for the coefficients of the last two terms to reflect the fact that these coefficients represent the difference in coefficient between 2026 and other years.

We found that neither interaction term proved to be statistically significant (see Table~\ref{tab:interact} for their p-values).
We can formally compare two models using the residual deviance. From Tables~\ref{tab:quarterly} and \ref{tab:interact}, we see that the residual deviance decreases by 4.5 between models while the number of residual degrees of freedom decreases by two. We can convert this to a p-value as follows:
\begin{align}
\text{p-value} &= 1 - \text{CDF}^{-1}_{\chi^2}(4.5; 2) \simeq 0.1 \,,
\end{align}
where $\text{CDF}^{-1}_{\chi^2}$ is the inverse CDF of a chi-squared distribution (with, in this case, 2 degrees of freedom).
Once again, we fail to obtain evidence sufficient to prove that early 2026 differed from the norm in any way.

\begin{table}%[!b]
    \small\centering
    
    \begin{tabular}{l%
        S[table-format=2.2]
        S[table-format=1.2]c}
     & \multicolumn{1}{c}{estimate} 
     & \multicolumn{1}{c}{std.\ error} 
     & p-value \\ \hline
    $\beta_\text{Q1}$ & 4.28 & 0.15 & 1.2e$-$188 \\
    $\beta_\text{Q2}$ & 3.84 & 0.15 & 1.1e$-$144 \\
    $\beta_\text{Q3}$ & 4.23 & 0.15 & 1.2e$-$184 \\
    $\beta_\text{Q4}$ & 4.56 & 0.14 & 1.2e$-$220 \\
    $\beta_\text{rep}$ & -1.13 & 0.04 & 1.9e$-$197 \\
    $\beta_{2026}$ & -0.93 & 0.07 & 1.9e$-$001 \\
    $\beta_\text{rep, 2026}$ & 0.30 & 0.18 & 1.0e$-$001
    \\[5pt]
    \multicolumn{4}{l}{Residual deviance: 280.0} \\
    \multicolumn{4}{l}{Residual deg.\ freedom: 237} \\
    \end{tabular}

    \caption{Coefficient table and residual deviance and degrees of freedom obtained by fitting eq.\,\ref{eq:interact} to the AMS data.}
    \label{tab:interact}
\end{table}

\subsection{Monthly rates}
\label{sec:month}

In this section, we bin the data by month rather than quarter in order to test for month-to-month variations in event counts. Rather than assume a functional form for the monthly variations, we treat ``month'' as a categorical variable. This is similar to our treatment of quarter in \S\ref{sec:rev} and is equivalent to performing a separate, intercept-only fit for each month of the year:
\begin{align}
    \ln \mu_{n,\,m_k} =~&\beta_{m_k} + \beta_\text{rep} \ln x_\text{rep} + \ln \, ( x_\text{yr}'' - 2009 ) 
    \label{eq:monthly} \, .
\end{align}
Here $m_k \in \lbrace \text{Jan}, ... \, , \text{Dec} \rbrace$ is the month and $x_\text{yr}''$ the fractional year:
\begin{align}
    x_\text{yr}'' &= x_\text{yr} + \frac{k - 2}{12} \, ,
\end{align}
where $k$ is the month index.
Note that we have again set ${\beta_\text{yr} = 1}$. 

\begin{figure}%[!hb]
    
    \includegraphics{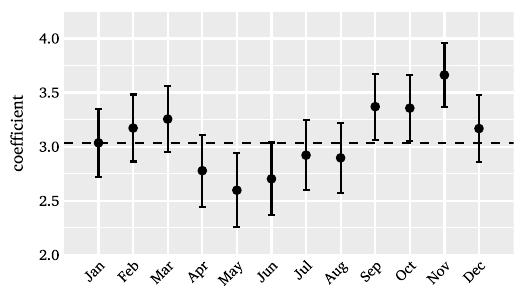}
    \captionof{figure}{Month-specific fit coefficient ($\beta_{m_k}$, black points) compared to the overall monthly average (horizontal dashed line). Error bars encompass the 95\% confidence interval for each month.}
    \label{fig:pattern}
\end{figure}

Figure~\ref{fig:pattern} shows the best-fitting value of $\beta_{m_k}$ for each month, along with its 95\% confidence interval; the coefficients are also provided in Table~\ref{tab:monthly}. There is fairly clear evidence that the rate varies from month to month, with May having the lowest number of reports and November the highest ($\beta_{m_k}$ is the log of the relative activity level). 
The number of reported fireballs in February, however, is consistent with the average.

\begin{table}
    \small\centering

    \begin{tabular}{l%
        S[table-format=2.2]
        S[table-format=1.2]c}
     & \multicolumn{1}{c}{estimate} 
     & \multicolumn{1}{c}{std.\ error} 
     & \multicolumn{1}{c}{p-value} \\ \hline
    $\beta_\text{Jan}$ & 3.04 & 0.16 & 1.5e$-$083 \\
    $\beta_\text{Feb}$ & 3.17 & 0.15 & 3.0e$-$093 \\
    $\beta_\text{Mar}$ & 3.26 & 0.15 & 4.1e$-$100 \\
    $\beta_\text{Apr}$ & 2.78 & 0.17 & 8.3e$-$063 \\
    $\beta_\text{May}$ & 2.60 & 0.17 & 5.3e$-$052 \\
    $\beta_\text{Jun}$ & 2.70 & 0.17 & 2.0e$-$058 \\
    $\beta_\text{Jul}$ & 2.92 & 0.16 & 5.7e$-$073 \\
    $\beta_\text{Aug}$ & 2.90 & 0.16 & 1.9e$-$071 \\
    $\beta_\text{Sep}$ & 3.37 & 0.15 & 2.6e$-$108 \\
    $\beta_\text{Oct}$ & 3.36 & 0.15 & 2.4e$-$107 \\
    $\beta_\text{Nov}$ & 3.66 & 0.15 & 5.1e$-$135 \\
    $\beta_\text{Dec}$ & 3.17 & 0.16 & 3.0e$-$092 \\
    $\beta_\text{rep}$ & -1.12 & 0.04 & 3.5e$-$202 
    \\[5pt]
    \multicolumn{4}{l}{Residual deviance: 917.0} \\
    \multicolumn{4}{l}{Residual deg.\ freedom: 719} \\
    \end{tabular}

    \captionof{table}{Coefficient table and residual deviance and degrees of freedom obtained by fitting eq.\,\ref{eq:monthly} to the AMS data.}
    \label{tab:monthly}
\end{table}

\begin{figure}
    \centering
    \includegraphics{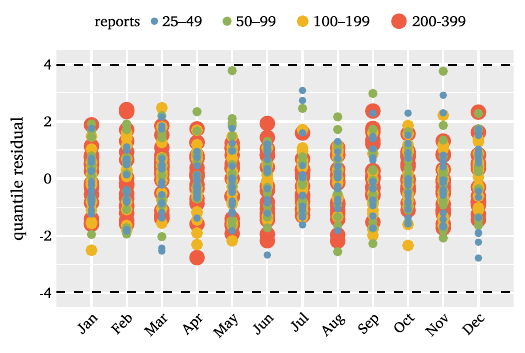}
    \caption{Quantile residuals for the number of fireballs reported to the AMS in each month, year, and reporting interval. Residuals are calculated relative to the best fit to eq.\,\ref{eq:monthly}. The dashed black lines at ${\pm 3.98}$ correspond to a Bonferroni-adjusted significance threshold of $\alpha_\text{test} = 0.5/732$. There are no outliers present.}
    \label{fig:mout}
\end{figure}

Now that we have fit a model that allows for annually recurring monthly variations, we can examine the residuals (Fig.\,\ref{fig:mout}) to determine if any month is a clear outlier. We notice that there are quite a few residuals with absolute values exceeding two (although these do not include any months in the first quarter). Therefore we also include dashed lines at ${\pm 3.98}$, which corresponds to ${\alpha_\text{test} = 0.05/732 = 6.8 \times 10^{-5}}$. No residuals exceed this Bonferroni-adjusted cutoff: in other words, the data do not show an anomalously high (or low) number of reported fireballs in any month -- including March of 2026 -- in the past 16 years at any reporting threshold (claim \#3).

\subsection{Events with delayed sound}
\label{sec:sound}

To investigate whether the number of events with reports of delayed sound was unusually high in the first quarter of 2026 (claim \#5), we performed a test for a difference in this proportion within each of our four reporting threshold bins. As the number of events in some bins is small, we use Fisher's exact test for a difference in proportions \citep{fisher22,agresti13} rather than a chi-squared test. The resulting p-values -- along with the fraction of events with delayed sound reports in each time period -- are presented in Table~\ref{tab:prop}.

\begin{table}[b]
    \centering\small
    \begin{tabular}{cccc} \hline \hline
        reports & before 2026 & in 2026 & p-value \\ \hline
        25--49 & 0.39 & 0.23 & 0.13	 \\
        50--99 & 0.61 & 0.75 & 0.20	\\
        100--199 & 0.62 & 0.82 & 0.22 \\
        200--399 & 0.66 & 0.75 & 1.00 \\	
    \end{tabular}
    \caption{Results of a test for a difference in the proportion of events with delayed sound within each reporting threshold bin. We include the fraction of events with sound prior to and during 2026 as well as the p-value for Fisher's  exact test.}
    \label{tab:prop}
\end{table}

Because we performed a set of four tests, we should adjust our significance threshold by a factor of four: ${\alpha_\text{test} = 0.05/4}$. A p-value smaller than this threshold would be evidence of a change in proportion. However, none of our p-values lie below the nominal significance level of 0.05, much less the adjusted value, and so we conclude that there is insufficient evidence of a change in the proportion of events with delayed sound.

\subsection{Radiant clustering}
\label{sec:rad}

\begin{figure*}%[b]
    \centering
    \includegraphics{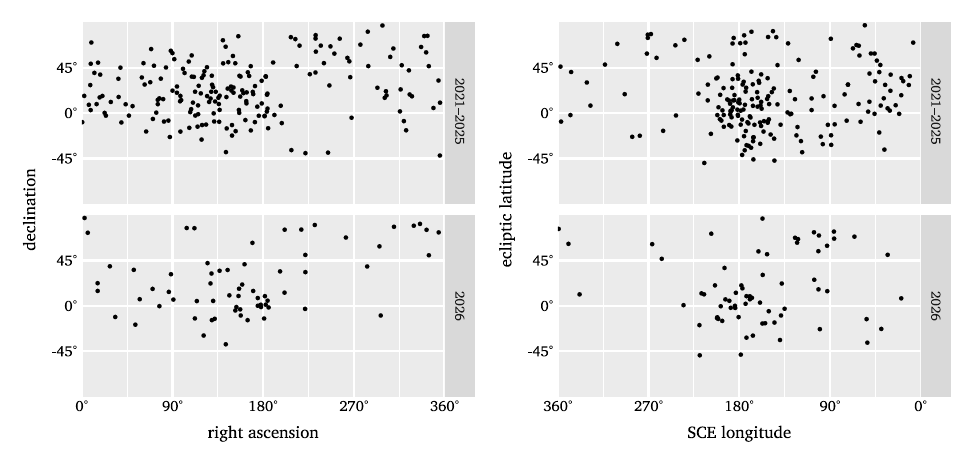}
    \caption{Radiants of fireballs reported to the AMS, separated by year as labeled and shown in both equatorial (left) and Sun-centered ecliptic (SCE; right) coordinates.}
    \label{fig:rad}
\end{figure*}

Finally, we examined the radiants of first-quarter events with at least 25 reports during the 2021--2026 period \citep[the same period used by][]{hankey26}. We used the dates -- supplemented by the times given in the downloaded data -- and equatorial radiants provided by \cite{hankey26} to calculate the solar longitude and SCE radiant for each event. The radiants in both coordinate systems are shown in Fig.\,\ref{fig:rad}. 
To our eye, the radiant distribution in 2026 is not obviously different from that in previous years.

\cite{hankey26} claims a difference in the number of reported fireballs within certain radiant ranges (claim \#7). Because there are an infinite number of ways to subset radiant space, we prefer to take a different approach and test for a difference in the overall distribution. We performed a two-sample, two-dimensional Kolmogorov-Smirnov (K-S) test \citep{fasano87} on the radiant distribution. Because spherical coordinates ``wrap'' around a sphere, we centered the data on ${(180^\circ, \, 0^\circ)}$, as shown in Fig.\,\ref{fig:rad}, so that we are cutting the SCE radiant distribution at a longitude where there are relatively few data points. 
If the K-S test returns a small p-value, that would be evidence of a change in distribution. In contrast, we obtained a p-value of 0.7418 when using equatorial radiants and 0.7678 when using SCE radiants, indicating a lack of evidence for a change in the radiant distribution between 2021--2025 and 2026.

\cite{hankey26} notes that these radiants are uncertain, and estimates the uncertainty to be $10^\circ$--$20^\circ$. This is comparable to the accuracy reported by \cite{moser17} for eyewitness-derived trajectories, albeit using a different methodology.
However, we compared radiants obtained through the API with those in the radiant file provided by \cite{hankey26}, and noticed that many had large discrepancies. Of the 257 events we compared, approximately one-third (92) had radiant discrepancies of more than $90^\circ$. For many events, the listed number of reports was also discrepant between the two files, suggesting that the radiant was recalculated after more reports were associated with an event. This did not appear to be the main cause of the radiant discrepancy, however; when we restricted our comparisons to events whose report total was consistent between data files, we found that 41 out of 131 events still had radiant discrepancies exceeding $90^\circ$. We therefore suspect that many events have radiant uncertainties much larger than $90^\circ$.

Finally, it is not clear whether these radiants are apparent or geocentric \citep[i.e., corrected for Earth's gravity;][]{gural01}. The difference between the two can be tens of degrees for slow-moving meteors on shallow entry angles.

\subsection{Untested claims}
\label{sec:untested}

We did not test two of the claims made in \cite{hankey26}. First, we were not able to test whether long-duration meteors are more numerous in 2026 (claim \#6) because duration information was not readily available from the AMS post or website. Duration information was available through the API, but we were unsure how much confidence to place in this data given the radiant discrepancies observed.

We also did not test any claims regarding ``major'' fireballs (claim \#4), as the definition of ``major'' was unclear to us. The list provided by \cite{hankey26} appears to include both very large and double events.

\newpage
\section{Conclusions}

We were able to test five out of seven claims made by the American Meteor Society regarding reported fireball events in the first quarter of 2026. Preserving the numbering used in the introduction of this paper, our findings are as follows:
\begin{enumerate}[itemsep=0em]
    \item[1.] The number of fireballs reported to the AMS appears to be increasing linearly with time. The number of fireballs seen in the first quarter of 2026 is consistent with this trend (\S\ref{sec:rev}).

    \item[2.] We did not find evidence of a change in the distribution of ``size'' (number of reports) of fireballs reported to the AMS in the first quarter of 2026 (\S\ref{sec:interact}).

    \item[3.] The number of fireballs reported in March 2026 is consistent with overall trends in fireball reports (i.e., linear growth over time and month-to-month variations; \S\ref{sec:month}).
    
    \item[5.] We did not find evidence of a change in the number of first-quarter events with delayed sound in 2026 (\S\ref{sec:sound}).
    
    \item[7.] We did not find evidence of a change in the radiant distribution of fireballs reported in the first quarter of 2026 vs.\ the first quarter of the years 2021--2025 (\S\ref{sec:rad}).

\end{enumerate}
The two remaining claims (\#4 and \#6) from \cite{hankey26} could not be tested due to either ambiguity or lack of data.

Although the number of reported fireballs does vary with time of year, the data did not support the ``February fireballs'' phenomenon. The rate at which fireballs are reported to the AMS in February is quite similar to the average monthly rate, and there is also little evidence of a change in the size (brightness) distribution. The month with the highest rates of reported fireballs is November. This is likely influenced by factors such as weather, hours of nighttime at the observers' locations, and human activity patterns; an exploration of these factors is beyond the scope of this study.

In section~\ref{sec:multiple}, we stressed the importance of accounting for the number of tests conducted when evaluating the significance of test results. We used this approach to assess the number of outliers (none) present in the quarterly and monthly rates. We also noted that one should ideally account for the total number of \emph{all} tests performed, rather than just the number of tests in a group. Since we have tested five hypotheses in total, this could be accomplished by dividing the per-test significance threshold, $\alpha_\text{test}$, by an additional factor of five. We do not take this additional step due to the lack of significant test results: if $p > \alpha_\text{test}$, then we can conclude that $p > \alpha_\text{test}/5$ as well.

We would like to emphasize that lack of evidence does \emph{not} mean that an effect does not exist: it means only that we cannot prove its existence from the AMS data currently available to us. Fortunately, reports to the AMS appear to be on the rise; if this pattern continues, non-random variations in rates will become easier to detect. However, statistically significant variations could arise as a result of unusually good or bad weather, or perhaps due to anthropogenic factors such as media influence. We anticipate that separating these factors from true variations in fireball rates will be extremely challenging.

Finally, variation in reported fireball rates need not be statistically significant to have real-world consequences. NASA's Meteoroid Environment Office (MEO) routinely analyzes bright fireballs of public interest, including fireballs with more than 75 reports to the AMS. The number of first-quarter fireballs in this category has roughly doubled between 2025 and 2026; this means that twice as many person-hours have been spent on fireball analyses this year. Characterizing the growth in mean fireball reporting rates and the variation about this mean will enable the MEO to better allocate resources in future years.

\newpage
\bibliographystyle{cas-model2-names}
\RaggedRight\small
\setlength{\bibsep}{0pt plus 0.3ex}
\bibliography{refs}

\end{document}